\documentclass{ws-ijgmmp}

\begin{document}

\markboth{G Marmo, G Scolarici, A Simoni, F Ventriglia } {
Alternative algebraic structures from bi-Hamiltonian quantum
systems}

%
\catchline{}{}{}{}{}
%

\title{ALTERNATIVE ALGEBRAIC STRUCTURES FROM BI-HAMILTONIAN QUANTUM SYSTEMS }

\author{G. MARMO\footnote{Dip. Sc. Fisiche, Universit\'a Federico II,
Via Cintia 80126 Napoli, Italy.}}

\address{Dip. Sc. Fisiche and INFN-Napoli, Universit\'a Federico II, Via Cintia \\
Napoli, 80126, Italy
\\
\email{marmo@na.infn.it}}

\author{G. Scolarici}

\address{Dip. Fisica and INFN-Lecce, Universit\'a di Lecce, Via per Arnesano\\
Lecce, 73100, Italy
\\
\email{scolarici@le.infn.it}}

\author{A. Simoni}

\address{Dip. Sc. Fisiche and INFN-Napoli, Universit\'a Federico II, Via Cintia \\
Napoli, 80126, Italy
\\
\email{simoni@na.infn.it}}

\author{F. Ventriglia}

\address{Dip. Sc. Fisiche and INFN-Napoli, Universit\'a Federico II, Via Cintia \\
Napoli, 80126, Italy
\\
\email{ventriglia@na.infn.it}}

\maketitle

\begin{history}
\received{(Day Month Year)} \revised{(Day Month Year)}
\end{history}

\begin{abstract}
We discuss the alternative algebraic structures on the manifold of
quantum states arising from alternative Hermitian structures
associated with quantum bi-Hamiltonian systems. We also consider the
consequences at the level of the Heisenberg picture in terms of
deformations of the associative product on the space of observables.
\end{abstract}

\section{Introduction}

Complete integrability in the classical framework has been associated with
the existence of alternative Hamiltonian descriptions. As a matter of fact,
integrability of a given dynamical system admitting alternative Hamiltonian
descriptions is usually implied by the compatibility of the two Hamiltonian
structures, when they are generic.

Because quantum theory is considered to be more fundamental than the
classical one, one usually expects that quantum systems, whose
corresponding classical limits are completely integrable, should
admit alternative quantum descriptions. Recently, some alternative
``quantum structures'' have been identified as alternative Hermitian
structures on the vector space of physical states in the
Schroedinger picture,\cite{morandi, ventriglia, romp, nagi1, lecce,
gallipoli, bedlevo, naymJPA} or as alternative associative products
on the space of observables in the Heisenberg picture.\cite
{carinena} In some of previous papers the notion of mutually
compatible alternative structures has been analyzed to find out how
many different dynamical systems may be bi-Hamiltonian with respect
to two given structures.

It is usually stated that the Schroedinger picture and the Heisenberg
picture are equivalent. In this paper, we would like to consider those
alternative structures in the Heisenberg picture which correspond to the
alternative Hermitian structures which we find in the Schroedinger picture.
More generally, when a dynamical system is identified with a derivation of
suitable products, we would like to consider the existence of alternative
and mutually compatible algebraic structures on the same carrier space. In
some sense, we use the notion of compatibility for Poisson brackets as a
guiding idea to define a notion of compatibility for algebraic structures
admitting a common derivation.

Moreover, a systematic formulation of quantum mechanics on
quaternionic Hilbert spaces exists \cite{adl} and, in the last few
years, there have been some suggestions that quaternionic quantum
mechanics (QQM) may be useful to classify positive maps in standard
quantum mechanics.\cite{kossa} Thus, we introduce the notion of
compatible quaternionic quantum bi-Hamiltonian dynamical systems and
discuss their Schroedinger and Heisenberg representations.

It is possible a discussion of both the standard and quaternionic
formulation of quantum mechanics in a unified conceptual framework
arising from a geometrization of quantum mechanics. Roughly
speaking, it is possible to start with a real differential manifold
$\mathcal{M}$ as carrier space,
instead of the Hilbert space $\mathcal{H}$. Then operators acting on $%
\mathcal{H}$ may be associated with (1-1)-tensors acting on the
tangent space $T\mathcal{M}$, and the coefficients of such tensor
fields at each point may be in turn real, or complex or even
quaternionic numbers. In other words, only the coefficients of
(1-1)-tensor fields are complex valued or quaternionic valued
functions, while the carrier space $\mathcal{M}$\ remains a real
differential manifold.

\section{Geometrization of Quantum Mechanics}

In this section we discuss the relevant geometric structures which
appear in standard quantum mechanics and the relations among them,
in the framework of our geometrization.\cite{bedlevo, geom}

To avoid technicalities, for the time being, while we deal with
general aspects, we shall confine ourselves to finite dimensional
carrier spaces.

We start with a complex Hilbert space $\mathcal{H}$ and consider its
realification $\mathcal{H}^{\mathbb{R}}.$ In other words, given a basis \{$%
\varphi _{k}$\}, any vector $\psi $ will be replaced by its complex
components $(q_{k}+ip_{k})$ in $\mathcal{H}$ and with real components $%
(q_{k},p_{k})$ in $\mathcal{H}^{\mathbb{R}}$. Now let the real vector space $%
\mathcal{H}^{\mathbb{R}},$ considered as a contractible real manifold $%
\mathcal{M,}$ be equipped with a symplectic structure $\omega ,$ a
non-degenerate 2-form such that
\begin{equation}
d\omega =0
\end{equation}
Then $\dim \mathcal{M}$ is even, say $2n.$ A global Darboux chart $\left\{
q_{k},p_{k}\right\} $ endows $\mathcal{M}$ with a real linear structure $%
\Delta ,$the infinitesimal generator of dilation (often also called the
Liouville vector field or the Euler operator) whose tensorial expression is
provided by
\begin{equation}
\Delta =\sum\nolimits_{k}\left( q_{k}\frac{\partial }{\partial q_{k}}+p_{k}%
\frac{\partial }{\partial p_{k}}\right) .
\end{equation}

In this chart, we define a linearly admissible complex structure, i.e. a
(1-1) tensor field $J$\ commuting with $\Delta $ such that $J^{2}=-1$. Then
construct a tensor\thinspace $g$ as
\begin{equation}
g=\omega \circ J.
\end{equation}
The triple $\left( g,\omega ,J\right) $ is (linearly) admissible if
$g$ results an Euclidean metric tensor in a global Darboux chart.
This generalizes the definition of admissible triple $\left(
g,\omega ,J\right) $ we have given in Ref. \cite{morandi}. It is
also possible to construct a (linearly) admissible triple $\left(
g,\omega ,J\right) $ starting from $g$ and $J,$ following the lines
of Ref. \cite{ruggiero}.

Along with a symplectic structure, an associated Poisson structure
$\Lambda =\omega ^{-1}$\ may be defined in the chosen global Darboux
chart as the contravariant counterpart of $\omega $, it corresponds
to the standard Poisson Brackets associated with a symplectic
structure.

To completely turn entities depending on the vector space structure on the
space of states into tensorial objects, we notice that with every matrix $%
A\equiv \left\| A_{k}^{j}\right\| \in \mathfrak{gl}(2n,\mathbb{R})$
acting on $\mathcal{H}^{\mathbb{R}}$ we can associate both a linear
vector field, acting on $\mathcal{M}:$
\begin{equation}
X_{A}:\mathcal{M}\rightarrow T\mathcal{M},\quad \psi \rightarrow \left( \psi
,A\psi \right)
\end{equation}
and a $(1-1)-$tensor, acting on $T\mathcal{M}:$%
\begin{equation}
T_{A}:T\mathcal{M}\rightarrow T\mathcal{M},\quad \left( \psi ,\varphi _{\psi
}\right) \rightarrow \left( \psi ,A\varphi _{\psi }\right) .
\end{equation}
So, when $A=1,$ we get the linear structure $\Delta :$
\begin{equation}
\Delta :\psi \rightarrow (\psi ,\psi ).
\end{equation}
This vector field allows to identify $T_{\psi }\mathcal{M}$ with $\mathcal{M}
$, i.e. the base manifold $\mathcal{M}$ gets a vector space structure from
the one on its tangent space at the origin.

Then, $X_{A}$ and $T_{A}$ are connected by the linear structure $\Delta :$%
\begin{equation}
T_{A}(\Delta )=X_{A}  \label{identify}
\end{equation}
and are both homogeneous of degree zero, i.e.
\begin{equation}
L_{\Delta }X_{A}=L_{\Delta }T_{A}=0.
\end{equation}
While the correspondence $A\rightarrow T_{A}$ is a full associative
algebra and a corresponding Lie algebra isomorphism, the correspondence $%
A\rightarrow X_{A}$ is instead only a Lie algebra (anti)isomorphism,
that is
\begin{equation}
T_{A}\circ T_{B}=T_{AB}
\end{equation}
while
\begin{equation}
\lbrack X_{A},X_{B}]=-X_{[A,B]}.
\end{equation}
Moreover, for any $A,B\in \mathfrak{gl}(2n,\mathbb{R}):$
\begin{equation}
L_{X_{A}}T_{B}=-X_{[A,B]}.  \label{tensorcost}
\end{equation}
Equation (\ref{tensorcost}) allows for the definition of constant tensors: a tensor $%
T_{B}$ is constant with respect to the linear structure $\Delta $ when
\begin{equation}
L_{\Delta }T_{B}=0.
\end{equation}

We recall that an Hermitian tensor $h$\ on $\mathcal{M},$ can be defined as
a map:
\begin{equation}
h:T_{\psi }\mathcal{M}\times T_{\psi }\mathcal{M}\longrightarrow \mathbb{C},
\label{targhet}
\end{equation}
such that
\begin{equation}
h(\Delta ,\Delta )=g(\Delta ,\Delta );\quad h(\Delta ,J(\Delta ))=i\omega
(\Delta ,J(\Delta ))
\end{equation}
are respectively a real valued and a purely imaginary valued quadratic
function of real variables.

We shall use the real quadratic function
\begin{equation}
\mathfrak{g}\mathbb{=}\frac{1}{2}g(\Delta ,\Delta )  \label{met}
\end{equation}
as the Hamiltonian generating function of the field $\Gamma $ :
\begin{equation}
i_{\Gamma }\omega =-d\mathfrak{g.}.  \label{gamma}
\end{equation}
It would have been possible to start with $J$ and $g$\ to recover
$\omega $ by means of the exterior derivative associated with
$J.$\cite{InverseMorandi} Indeed, with the help of
\begin{equation}
d_{J}=d\circ J-J\circ d,
\end{equation}
the symplectic structure is recovered through
\begin{equation}
dd_{J}\left( \frac{1}{2}g(\Delta ,\Delta )\right) =\omega .
\end{equation}

It is possible to show that
\begin{equation}
\Gamma =J(\Delta )  \label{jdelta}
\end{equation}
and $J(\Gamma )=-\Delta $. The vector field $\Gamma $ preserves all three
structures $g,\omega $ and $J$. Thus the vector field $\Gamma $ will be a
generator of a one-parameter group of unitary transformations and may be
associated with a Schroedinger-type equation (we set $\hslash =1$)
\begin{equation}
J\frac{d}{dt}\psi =H\psi ,
\end{equation}
where $H\psi $\ is the second component of $\Gamma (\psi ).$\ The dynamics
is therefore determined by the vector field $\Gamma $.

To search for alternative descriptions, we look for all Hermitian tensors on
$\mathcal{M}$ invariant under the dynamical evolution. We have to consider
the equation $L_{\Gamma }h=0$ for the unknown $h,$ $h$ representing an
unknown Hermitian tensor on $\mathcal{M}$. This is equivalent to $L_{\Gamma
}\omega =0$, $L_{\Gamma }g=0$, $L_{\Gamma }J=0$, so that we may solve for $%
L_{\Gamma }h=0$ by starting solving for $L_{\Gamma }\omega =0$. In
this way we take into account, as discussed in Ref. \cite{dub}, that
both $\omega $ and $g$ may be point-dependent. Here, rather than
dealing with the general theory, we limit ourselves to discuss a
simple example.

\subsection{A simple example}

For a one-dimensional system, in a global Darboux chart $\left( q,p\right) $
of $\mathcal{M=}\mathbb{R}^{2}$, we consider the dynamics described by the
vector field
\begin{equation}
\Gamma =p\frac{\partial }{\partial q}-q\frac{\partial }{\partial p}
\end{equation}
with standard Hamiltonian description provided by
\begin{equation}
\omega _{s}=dq\wedge dp,\,\,\,\ H_{s}=\frac{1}{2}(q^{2}+p^{2}).
\end{equation}
The other relevant tensors are
\begin{equation}
g=dq\otimes dq+dp\otimes dp,
\end{equation}
\begin{equation}
J=dp\otimes \frac{\partial }{\partial q}-dq\otimes \frac{\partial }{\partial
p},
\end{equation}
\begin{equation}
\Delta =q\frac{\partial }{\partial q}+p\frac{\partial }{\partial p}
\end{equation}
and
\begin{equation}
\Lambda =\frac{\partial }{\partial p}\wedge \frac{\partial }{\partial q}.
\end{equation}

The most general symplectic structure solving the previously stated
equation for $\omega $ is given by \cite{romp 1}
\begin{equation}
\omega _{F}=F(H_{s})dq\wedge dp
\end{equation}
with $F(H_{s})$ vanishing nowhere.

By performing simple computations, one finds for every $\omega _{F}$ a
Darboux chart
\begin{equation}
P=p(1+f(H_{s})),\,\,\,\ Q=q(1+f(H_{s})),
\end{equation}
where $f$ is any solution of the differential equation
\begin{equation}
\frac{d}{ds}[s(1+f(s))^{2}]=F(s).
\end{equation}
As a particular case, we may consider a nonlinear diffeomorphism of the form
\begin{equation}
P_{\lambda }=p(1+\lambda H_{s}),\ \ Q_{\lambda }=q(1+\lambda H_{s}),
\label{nltr}
\end{equation}
where a real parameter $\lambda $ appears. From Eq.(\ref{nltr}) we get
\begin{equation}
dP_{\lambda }=(1+\lambda (q^{2}+3p^{2}))dp+2\lambda pqdq,\ \ dQ_{\lambda
}=(1+\lambda (p^{2}+3q^{2}))dq+2\lambda pqdp.  \label{differenziali}
\end{equation}
By using Eq.(\ref{differenziali}), the metric tensor $g_{\lambda }$ and the
symplectic form $\omega _{\lambda }$ can be respectively obtained as
functions of $q,p$ as
\begin{eqnarray}
g_{\lambda } &=&dP_{\lambda }\otimes dP_{\lambda }+dQ_{\lambda }\otimes
dQ_{\lambda }  \label{metrica} \\
&=&[(1+\lambda (q^{2}+3p^{2}))^{2}+4\lambda ^{2}p^{2}q^{2}]dp\otimes dp+
\nonumber \\
&&[(1+\lambda (p^{2}+3q^{2}))^{2}+4\lambda ^{2}p^{2}q^{2}]dq\otimes dq+
\nonumber\\
&&4\lambda pq(1+2\lambda (q^{2}+p^{2}))[dq\otimes dp+dp\otimes dq]
\nonumber
\end{eqnarray}
and
\begin{eqnarray}
\omega _{\lambda } &=&dP_{\lambda }\wedge dQ_{\lambda }  \label{symp} \\
&=&[(1+\lambda (q^{2}+3p^{2}))^{2}-4\lambda ^{2}p^{2}q^{2}]dp\wedge
dq. \nonumber
\end{eqnarray}
The associated Poisson bracket in the $\left( Q,P\right) $
coordinates will be given by the following expression
\begin{equation}
\{Q,P\}_{\lambda }=\frac{1}{[(1+\lambda (q^{2}+3p^{2}))^{2}-4\lambda
^{2}p^{2}q^{2}]}\{q,p\}  \label{poissonb}
\end{equation}
due to the use of a non-canonical transformation. As for the complex
structure we have
\begin{equation*}
J(p,q)=dp\otimes \frac{\partial }{\partial q}-dq\otimes \frac{\partial }{%
\partial p}.
\end{equation*}
In the $(Q,P)$ coordinates the dynamical
vector field $\Gamma $ has the form
\begin{equation}
\Gamma =P\frac{\partial }{\partial Q}-Q\frac{\partial }{\partial P}.
\end{equation}

We see immediately that $\Gamma $ is also Hamiltonian with respect to an
alternative Poisson bracket given by $\left\{ Q,P\right\} =1,$ along with
the complex structure
\begin{equation}
J(P,Q)=dP\otimes \frac{\partial }{\partial Q}-dQ\otimes \frac{\partial }{%
\partial P}.  \label{JQP}
\end{equation}

We remark that the two vector fields
\begin{equation}
\Delta (p,q)=q\frac{\partial }{\partial q}+p\frac{\partial }{\partial p}%
,\quad \Delta (P,Q)=Q\frac{\partial }{\partial Q}+P\frac{\partial }{\partial
P}
\end{equation}
define two alternative linear structures on $\mathcal{M,}$\ which are not
linearly related. Indeed the following tensor $T:$%
\begin{equation}
T=\frac{P}{p}\frac{\partial }{\partial P}\otimes dp+\frac{Q}{q}\frac{%
\partial }{\partial Q}\otimes dq,
\end{equation}
written for simplicity in mixed coordinates, maps one linear structure \`{i}%
nto the other:
\begin{equation}
T\left( \Delta (p,q)\right) =T\left( q\frac{\partial }{\partial q}+p\frac{%
\partial }{\partial p}\right) =Q\frac{\partial }{\partial Q}+P\frac{\partial
}{\partial P}=\Delta (P,Q).
\end{equation}

The existence of these two alternative linear structures means that
we may compose (add) solutions for $\Gamma $ in alternative ways to
get new solutions. The fact that these linear structures are not
linearly related means that the image of the composition (sum) is
not the composition (sum) of the images.\cite{Wignerproblem}

\section{Alternative compatible Hermitian structures}

In general, two symplectic structures associated with a classical
bi-Hamiltonian system do not admit a common Darboux chart. When they are
``constant'' in the same global Darboux chart, one can find a linear
transformation (that is, a diffeomorphism commuting with $\Delta $) which
connects the two symplectic structures.

In this section we review some results on quantum bi-Hamiltonian systems
concerning tensor fields which are compatible with (i.e. constant with
respect to) a given linear structure $\Delta $ and compatible with (i.e.
commuting with) an assigned complex structure\thinspace $J.$

Suppose that two admissible triples $(g_{1},J_{1},\omega _{1})$ and $%
(g_{2},J_{2},\omega _{2})$ are given on $\mathcal{M=H}^{\mathbb{R}}$. Then,
by complexification, we get two different Hilbert spaces, each one with its
proper multiplication by complex numbers and with its proper Hermitian
structure. Quantum theory in the usual Schroedinger formulation, when we
start from a given Schroedinger dynamics, leads quite naturally to consider
identical complex structures in the two triples and the condition $%
J_{1}=J_{2}$ is a sufficient condition for compatibility \cite{morandi}. On
the contrary, in the real context it is possible to consider the case of two
admissible triples with $J_{1}\neq J_{2}$ which are compatible \cite{naymJPA}%
.

Then, we may assume compatibility, by taking two Hermitian structures, $%
h_{1}(.,.)$ and $h_{2}(.,.)$, on the same complex Hilbert space $\mathcal{H}$%
, coming from two admissible triples admitting $J_{1}=J_{2}$. We search for
the group of automorphism which leave both $h_{1}$ and $h_{2}$ invariant,
that is the bi-unitary transformations group.

By using the Riesz's theorem a bounded, positive operator $G$ may be
defined, which is self-adjoint both with respect to $h_{1}$ and $h_{2}$, as:
\begin{equation}
h_{2}(x,y)=h_{1}(Gx,y),\ \ \ \ \forall x,y\in \mathcal{H}.
\end{equation}

Moreover, any bi-unitary transformation $U$ must commute with $G$. Indeed:
\begin{equation*}
h_{1}(x,U^{\dagger
}GUy)=h_{1}(Ux,GUy)=h_{2}(Ux,Uy)=h_{2}(x,y)=h_{1}(Gx,y)=h_{1}(x,Gy)
\end{equation*}
and from this
\begin{equation}
U^{\dagger }GU=G\Leftrightarrow \lbrack G,U]=0.
\end{equation}
Therefore the group of bi-unitary transformations is contained in the
commutant $G^{\prime }$ of the operator $G$.

To visualize these transformations, let us consider the bi-unitary group of
transformations when $\mathcal{H}$ is finite-dimensional. In this case $G$
is diagonalizable and the two Hermitian structures result proportional in
each eigenspace of $G$ \emph{via} the eigenvalue. Then the group of
bi-unitary transformations is given by

\begin{equation}
U(d_{1})\times U(d_{2})\times ...\times U(d_{m}),\ \ \
d_{1}+d_{2}+...+d_{m}=n=\dim \mathcal{H},
\end{equation}
where $d_{k}$ denotes the degeneracy of the $k$-th eigenvalue of $G$.

Each Hermitian structure on $\mathcal{H}$ defines a different realization of
the unitary group as a group of transformations. The intersection of these
two groups identifies the group of bi-unitary transformations.

Now we can further qualify and strengthen the compatibility condition by
stating the following \cite{morandi}:

\noindent \textbf{Definition 1 }\textit{Two Hermitian forms are said to be
in generic relative position when the eigenvalues of }$G$\textit{\ are
non-degenerate.}

Then, if \ $h_{1}$ and $h_{2}$ are in generic position, the group of
bi-unitary transformations becomes

\begin{eqnarray*}
&&\underbrace{U(1)\times U(1)\times ...\times U(1)}.  \nonumber \\
&&\ \ \ \ \ \ \ \ \ \ n\ \ factors  \nonumber
\end{eqnarray*}

In other words, this means that $G$ generates a complete set of commuting
observables.

Moreover, the following proposition holds \cite{naymJPA}:

\noindent \textbf{Proposition 1} \textit{Two Hermitian forms are in generic
relative position if and only if their connecting operator }$G$\textit{\ is
cyclic}.

This shows that definition $(1)$ may be equivalently formulated as:

\noindent \textbf{Definition 2} \textit{Two Hermitian forms are said to be
in generic relative position when their connecting operator }$G$\textit{\ is
cyclic.}

The genericity condition can also be restated in a purely algebraic form as
follows \cite{morandi}:

\noindent \textbf{Definition 3} \textit{Two Hermitian forms are said to be
in generic relative position when }$G^{\prime \prime }=G^{\prime }$\textit{,
i.e. when the bi-commutant of }$G$ \textit{coincides with the commutant of} $%
G$.

Equivalence of definitions $(1)$, $(2)$, and $(3)$ is apparent and hold in
the finite as well as in the infinite-dimensional case \cite{naymJPA}.

\noindent \textbf{Remark }In a more abstract setting, bi-Hamiltonian
dynamical systems associated with systems $1$ and $2$ can be seen as the
infinitesimal generators of one -parameter groups in the intersection
\begin{equation*}
Aut\ S_{1}\cap Aut\ S_{2},
\end{equation*}
where $Aut\ S$ denotes the automorphisms of the structure $S$. In
classical and in quantum mechanics, $S_{1}$ and $S_{2}$ will
represent symplectic and Hermitian structures respectively
associated with the systems $1$ and $2$ and the intersection $Aut\
S_{1}\cap Aut\ S_{2}$ will be associated with all bi-Hamiltonian
dynamical systems. In this geometrical context the compatibility can
be restated as follows:

\noindent \textbf{Definition 4} \textit{Two structures }$S_{1}$ \textit{and}
$S_{2}$\textit{\ will be said compatible iff the intersection }$Aut\
S_{1}\cap Aut\ S_{2}$ \textit{is non void and non trivial.}

Moreover, the following definition qualify and strengthen the compatibility
condition:

\noindent \textbf{Definition 5} \textit{Two (compatible) structures
}$S_{1}$ \textit{and} $S_{2}$\textit{\ will be said to be in
relative generic position iff the intersection }$Aut\ S_{1}\cap Aut\
S_{2}$ \textit{is minimal and non trivial.}

Equivalence of definitions $(1)$, $(2)$, $(3)$ and $(5)$ is apparent.

\section{Alternative compatible quaternionic Hermitian structures}

Equations of motion in (right) quaternionic Hilbert space $\mathcal{H}^{%
\mathbb{Q}}$ are defined by the Schroedinger equation (we set $\hslash =1$)
\cite{adl}:

\begin{equation}
\frac{d}{dt}\psi =-\widetilde{H}\psi .
\end{equation}

The dynamics is determined by the linear operator $\widetilde{H}$. To search
for alternative Hermitian quaternionic descriptions, we look for all scalar
products on $\mathcal{H}^{\mathbb{Q}}$ invariant under the dynamical
evolution.

Along the lines of previous sections, if we define $\Gamma :\mathcal{H}^{%
\mathbb{Q}}\rightarrow T\mathcal{H}^{\mathbb{Q}}$ to be the map $\psi
\rightarrow (\psi ,-\widetilde{H}\psi )$, we have to solve for $L_{\Gamma
}h=0$, where now $h$ is an unknown Hermitian quaternionic structure on $%
\mathcal{H}^{\mathbb{Q}}$.

We recall \cite{sh} that any $h$ on $\mathcal{H}^{\mathbb{Q}}$ defines an
Euclidean metric $g$, three symplectic forms $\omega _{a}$ $(a=1,2,3)$ and
three complex structures $J_{a}$ satisfying the quaternionic algebra on the
realification $\mathcal{H}^{\mathbb{R}}$ of the right quaternionic Hilbert
space $\mathcal{H}^{\mathbb{Q}}$:
\begin{equation}
h(.,.)=:g(.,.)+ig(J_{1}.,.)+jg(J_{2}.,.)+kg(J_{3}.,.)
\end{equation}
where $i,$ $j,$ $k$ are the quaternion units satisfying $ij=-ji=k$, $%
i^{2}=j^{2}=k^{2}=-1$.

The imaginary parts of $h$ are symplectic structures $\omega _{a}$ on the
real vector space $\mathcal{H}^{\mathbb{R}}$:
\begin{equation}
\omega _{a}(.,.):=g(J_{a}.,.).
\end{equation}
In the quaternionic case we define \emph{admissible }triple by $(g,\mathbf{J}%
,\mathbf{\omega })$, where $\mathbf{J}=(J_{1},J_{2},J_{3})$ and $\mathbf{%
\omega }=(\omega _{1},\omega _{2},\omega _{3})$ define hypercomplex
and hypersymplectic structures respectively.\cite{gaeta}

Introducing now two right quaternionic Hermitian structures $h_{1}$ and $%
h_{2}$ on the real space $\mathcal{H}^{\mathbb{R}}$, coming from two
admissible triples $(g_{1},\mathbf{J}_{1},\mathbf{\omega }_{1})$ and $(g_{2},%
\mathbf{J}_{2},\mathbf{\omega }_{2})$, we will show that sufficient
condition for compatibility according with definition 4, is that the
hypercomplex structures $\mathbf{J}_{1}$ and $\mathbf{J}_{2}$ are the same,
up to a transformation of a right $SU(2)$ group,\ i. e. $h_{1}$ and $h_{2}$
are defined, up to an automorphism\ of $\mathbb{Q}$, on the same right
quaternionic Hilbert space $\mathcal{H}^{\mathbb{Q}}.$

To show this, we start resuming well known results about symmetry
transformations on right quaternionic Hilbert spaces.

Physical states in QQM are in one-to-one correspondence with unit rays of
the form $|\mathbf{\psi }\rangle =\{|\psi \rangle \theta \}$, with $|\psi
\rangle $ a unit normalized vector and $\theta $ a ``quaternionic phase'' of
magnitude unity. A symmetry operation $\mathcal{S}$ of the system is a
mapping of the unit rays $|\mathbf{\psi }\rangle $ into images $|\mathbf{%
\psi }^{\prime }\rangle $, which preserves all transition probabilities:
\begin{equation}
\mathcal{S}|\mathbf{\psi }\rangle =|\mathbf{\psi }^{\prime }\rangle
\label{symmetry}
\end{equation}
\begin{equation*}
|\langle \mathbf{\psi }^{\prime }|\mathbf{\varphi }^{\prime }\rangle
|=|\langle \mathbf{\psi }|\mathbf{\varphi }\rangle |.
\end{equation*}

In\ CQM case, the classical Wigner theorem states that the unit ray mapping
of the previous equation can be replaced, by an appropriate choice of ray
representatives, by a mapping $\mathcal{U}|\psi \rangle =|\psi ^{\prime
}\rangle $ acting on the vectors $|\psi \rangle $ of Hilbert space, with $%
\mathcal{U}$ either unitary or antiunitary. This theorem was generalized by
Bargmann to the case of QQM \cite{bargman}. The generalized theorem states
that for a quantum mechanics based on a field $\mathbb{F}$, the unit ray
mapping of the previous equation can always be replaced by a vector mapping $%
\mathcal{U}|\psi \rangle =|\psi ^{\prime }\rangle $ where
$\mathcal{U}$ denotes an additive projective unitary transformation
obeying
\begin{eqnarray}
\mathcal{U}(|\psi \rangle +|\varphi \rangle ) &=&\mathcal{U}|\psi \rangle +%
\mathcal{U}|\varphi \rangle  \label{counitary} \\
\mathcal{U}|\psi q\rangle &=&\mathcal{U}|\psi \rangle Aut_{\mathcal{U}}(q)
\nonumber \\
h(\mathcal{U}\psi ,\mathcal{U}\varphi ) &=&Aut_{\mathcal{U}}(h(\psi
,\varphi )),  \nonumber
\end{eqnarray}
with $Aut_{\mathcal{U}}(q)$ a $\mathcal{U}$-dependent automorphism of the
field $\mathbb{F}$. When $\mathbb{F}$ is the field of quaternions $\mathbb{Q}
$, the automorphism $Aut_{\mathcal{U}}$ must have the form
\begin{equation*}
Aut_{\mathcal{U}}(q)=\overline{\theta }_{\mathcal{U}}q\theta _{\mathcal{U}},%
 \,\,\,\,\ |\theta _{\mathcal{U}}|=1
\end{equation*}
where $\overline{\theta }$ denotes the quaternion conjugate of $\theta $.

Defining now a new operator $U$ by
\begin{equation*}
U|\psi \rangle =\mathcal{U}|\psi \rangle \overline{\theta }_{\mathcal{U}}
\end{equation*}
for arbitrary $|\psi \rangle $ we immediately obtain from Eq.(\ref{counitary}%
) that
\begin{eqnarray*}
U|\psi q\rangle &=&U|\psi \rangle q \\
h(U\psi ,U\varphi ) &=&h(\psi ,\varphi )
\end{eqnarray*}
and so $U$ gives a quaternion linear, unitary vector mapping. Then, in QQM,
the unit ray mapping of Eq. (\ref{symmetry}) can always be replaced by a
unitary mapping $U$ on the same Hilbert space. With this fact in mind, we
now search for the unitary transformations which leave both $h_{1}$ and $%
h_{2}$ invariant, that is the bi-unitary transformations group. This group
of transformations allows us to construct the group of automorphisms of both
$h_{1}$ and $h_{2}$.

By using the Riesz's theorem (that also holds for right quaternionic Hilbert
spaces \cite{hor}) a bounded, positive operator $G$ may be defined, which is
self-adjoint both with respect to $h_{1}$ and $h_{2}$, as:
\begin{equation}
h_{2}(x,y)=h_{1}(Gx,y),\ \ \ \ \forall x,y\in \mathcal{H}^{\mathbb{Q}}.
\end{equation}

Moreover, as in the case of complex Hilbert space, any bi-unitary
transformation $U$ must commute with $G$:
\begin{equation}
U^{\dagger }GU=G\Leftrightarrow \lbrack G,U]=0.
\end{equation}
Therefore the quaternionic group of bi-unitary transformations is contained
in the commutant $G^{\prime }$ of the operator $G$. In the case that $G$
admits discrete spectrum, its spectral decomposition reads \cite{adl}

\begin{equation*}
G=\sum_{m}\sum_{a=1}^{d_{m}}|u_{m},a\rangle \lambda _{m}\langle u_{m},a|,%
\,\,\,\ \lambda _{m}>0,
\end{equation*}
where $\{|u_{m},a\rangle \}$ is the eigenvectors basis of $G$
(orthonormal with respect to both the Hermitian structures) and $a$
is a degeneracy label. The quaternionic commutant $U$ of the
operator $G$ reads (see proposition 4 of Ref. \cite{scola} )
\begin{equation*}
U=\sum_{m}\sum_{a=1}^{d_{m}}\sum_{b=1}^{d_{m}}|u_{m},a\rangle
u(m,a,b)\langle u_{m},b|,\,\,\,\,\ u(m,a,b)\in \mathbb{Q}.
\end{equation*}

Moreover $U$ is unitary if the square matrices $[u(m,a,b)]$ of dimension $%
d_{m}$\ with entries $u(m,a,b)$ ($a$ and $b$ denote row and column
indices respectively) belong to the quaternionic unitary group
$U(d_{m},\mathbb{Q})$ of dimension $d_{m}$.

Then, the quaternionic group of bi-unitary transformations is given by
\begin{equation}
U(d_{1},\mathbb{Q})\times U((d_{2},\mathbb{Q})\times ...\times U(d_{m},%
\mathbb{Q}),\ \ \ d_{1}+d_{2}+...+d_{m}=n.
\end{equation}

According with definitions $(1)$, if \ $h_{1}$ and $h_{2}$ are in generic
position, the group of\ bi-unitary transformations becomes
\begin{eqnarray}
&&\underbrace{U(1,\mathbb{Q})\times U(1,\mathbb{Q})\times ...\times U(1,%
\mathbb{Q})}.  \label{groupquatgen} \\
&&\ \ \ \ \ \ \ \ \ \ \ \ \ \ \ \ \ \ \ n\ \ factors  \nonumber
\end{eqnarray}

Now, we say that a quaternionic operator $G$ is cyclic when a vector $%
|x_{0}\rangle $ exists such that the set $\{|x_{0}\rangle ,$ $G|x_{0}\rangle
,...,$ $G^{n-1}|x_{0}\rangle \}$ spans the whole $n-$dimensional right
quaternionic vector space $\mathbb{Q}^{n}$, i.e. they are right linearly
independent on $\mathbb{Q}$,\ we show that:

\noindent \textbf{Proposition 2} \textit{Two quaternionic Hermitian forms
are in generic relative position if and only if their connecting operator }$%
G $\textit{\ is cyclic}.

\noindent \textbf{Proof\ }The non singular Hermitian operator $G$ has a
discrete spectrum and is diagonalizable so, when $h_{1}$ and $h_{2}$ are in
generic position, $G$ admits $n$ distinct real eigenvalues $\lambda _{l}$.
Let now $\{|u_{l}\rangle \}$ be the eigenvector basis of $G$ and $\{\mu
_{l}=c_{l}+jc_{l}^{\prime }\}$ an $n-$tuple of non-zero quaternionic
numbers. The vector
\begin{equation}
|x_{0}\rangle =\sum\nolimits_{l}|u_{l}\rangle \mu _{l}
\end{equation}
is a cyclic vector for $G$. In fact, by applying $G^{m}$ to $|x_{0}\rangle $
one obtains
\begin{equation}
G^{m}|x_{0}\rangle =\sum\nolimits_{l}|u_{l}\rangle \mu _{l}\lambda _{l}^{m}\
,\ \ \ m=0,1,...,n-1
\end{equation}
and the vectors $\{G^{m}|x_{0}\rangle \}$ are right linearly independent on $%
\mathbb{Q}$. In fact, it is known that the rank of a $n-$dimensional
quaternionic matrix is $n$ if and only if its complex counterpart\ has rank $%
2n.$\cite{zha} Then, denoting with $\Lambda $ the Vandermonde matrix
and with $C=diag(c_{1},\cdots ,c_{n})$ and $C^{\prime
}=diag(c_{1}^{\prime },\cdots ,c_{n}^{\prime })$ two diagonal
complex matrices, the complex counterpart\ of the quaternionic
matrix $M=\Lambda C+j\Lambda C^{\prime }$\ of the\ components is
given by
\begin{equation*}
M_{c}=\left(
\begin{array}{cc}
\Lambda C & \Lambda C^{\prime } \\
-(\Lambda C^{\prime })^{\ast } & (\Lambda C)^{\ast }
\end{array}
\right) =\left(
\begin{array}{cc}
\Lambda C & \Lambda C^{\prime } \\
-\Lambda C^{\prime \ast } & \Lambda C^{\ast }
\end{array}
\right) ,
\end{equation*}
and by a direct computation one has
\begin{equation}
\det M_{c}=(\prod\limits_{k}|\mu _{k}|^{2}V^{2}(\lambda _{1},...,\lambda
_{n}),
\end{equation}
where $V$ denotes the Vandermonde determinant which is different from zero
when all the eigenvalues $\lambda _{k}$ are distinct. The converse is also
true.$\ \ \Box $

The equivalence of definitions $(1)$, $(2)$ and $(4)$ is apparent\ also for
finite-dimensional right quaternionic vector spaces.

We conclude this Section noticing that, unlike the complex case, definition $%
(3)$ is not equivalent to definition $(1)$, $(2)$ and $(4)$ when two
Hermitian forms are considered on a right quaternionic vector space. In
fact, if for instance definition $(3)$ holds, it is easy to see that the
bi-commutant of $G$ is abelian while its commutant is not commutative
according with Eq.(\ref{groupquatgen}), hence $G^{\prime \prime }\neq
G^{\prime }$.

\section{Alternative compatible algebraic structures}

In this Section we will discuss the Heisenberg picture of Quantum
bi-Hamiltonian dynamical systems on complex and quaternionic Hilbert spaces
limiting ourselves to consider the case of constant tensorial structures.

Looking for alternative quantum Hamiltonian descriptions in the
Heisenberg picture is equivalent to search for alternative
associative products on the space of observables, with the
requirement that the equations of motion define a derivation with
respect to the alternative associative product.\cite {carinena}

We start with some pure algebraic considerations.\cite{carinena,
Wignerproblem}

Let $(\mathcal{A},\cdot )$ be a unital associative algebra with identity $E$%
. A simple way to define a new associative product on $\mathcal{A}$ is to
take an element $K\in $ $\mathcal{A}$ and to define a new product by
\begin{equation}
A\circ _{K}B=A\cdot K\cdot B.
\end{equation}
We denote with $(\mathcal{A},\circ _{K})$ the new associative algebra; if $K$
is invertible in $(\mathcal{A},\cdot )$, then $(\mathcal{A},\circ _{K})$ is
also a unital algebra. In this case, the identity in $(\mathcal{A},\circ
_{K})$ is $E_{K}=$ $K^{-1}$ and there is an isomorphism $\varphi $ between
the algebras $(\mathcal{A},\cdot )$ and $(\mathcal{A},\circ _{K})$:
\begin{equation}
\varphi :(\mathcal{A},\cdot )\rightarrow (\mathcal{A},\circ _{K}):\varphi
(A)=\frac{1}{K}\cdot A
\end{equation}

The two different associative products in $\mathcal{A}$ allow the
introduction of two different Lie products
\begin{equation}
\lbrack A,B]=A\cdot B-B\cdot A
\end{equation}
and
\begin{equation}
\lbrack A,B]_{K}=A\cdot K\cdot B-B\cdot K\cdot A=A\circ _{K}B-B\circ _{K}A.
\end{equation}
From now on we will skip the product symbol $\cdot $ for the original
associative structure.

Let us consider the following Lie subalgebras, $S\subset \mathcal{A}$%
\begin{equation}
S\subset \mathcal{A:\ }S=\{A:[A,K]=0\}
\end{equation}
and $S_{K}\subset \mathcal{A}$%
\begin{equation}
S_{K}\subset \mathcal{A:\ }S_{K}=\{A:[A,K]_{K}=0\}.
\end{equation}
It is immediate to verify that $S\subset S_{K}$; furthermore, if $K$ is
invertible (note that the invertibility notion does not depend on the
algebra considered) we can exchange the algebras so that it follows $S=S_{K}$%
.

In order to discuss now the correspondence between quantum bi-Hamiltonian
systems in the Schroedinger and in the Heisenberg pictures, we have to think
of $\mathcal{A}$ as an algebra of bounded operators on a Hilbert space, $%
\mathcal{H}$\ or $\mathcal{H}^{\mathbb{Q}}$, provided with two Hermitian
structures, $h_{2}(.,.)$ and $h_{1}(.,.)$. As said in sections 2 and 3, two
compatible (complex or quaternionic) Hermitian structures $h_{2}$ and $h_{1}$
are related by means of a positive self-adjoint operator $G=h_{1}^{-1}\circ
h_{2}$. Moreover, the presence of two Hermitian structures allows us to
define in $\mathcal{A}$ two involutions, the adjoints, denoted with $\dagger
$ and $\ast $ respectively. Then, the following relations hold
\begin{equation*}
h_{1}(Ax,By)=h_{1}(B^{\dagger }Ax,y)
\end{equation*}
and
\begin{equation}
h_{2}(Ax,By)=h_{2}(B^{\ast }Ax,y)=h_{1}(GAx,By)=h_{1}(B^{\dagger }GAx,y).
\label{adj}
\end{equation}

As a consequence of the fact that the adjoints operators $B^{\dagger }$ and $%
B^{\ast }$ belong to the algebra $\mathcal{A}$ and Eq.(\ref{adj}), it
follows that the alternative associative product between operators is given
by
\begin{equation}
A\circ _{G}B=AGB.  \label{associative}
\end{equation}

As said before, we obtain two operator algebras endowed with two different
associative products and two corresponding different Lie products. The
invertibility of $G$ leads to the isomorphism $S=S_{G}$. Furthermore, the
algebra $\ S=S_{G}$ is invariant with respect to the two involutions because
the operator $G$ is self-adjoint with respect to both the Hermitian
structures and the involutions coincide if we restrict ourselves to $S=S_{G}$%
.

What happens when $h_{2}$ and $h_{1}$ are in relative generic position on
complex or (right) quaternionic Hilbert spaces? \

If $h_{2}$ and $h_{1}$ are in relative generic position on
$\mathcal{H}$, the previous analysis shows that $S$ and $S_{G}$ are
two abelian isomorphic algebras. Moreover, these two algebras
decompose into a direct sum of one-dimensional complex algebras.

On the contrary, if $h_{2}$ and $h_{1}$ are not in relative generic position
on $\mathcal{H}$, the previous analysis shows that $S$ and $S_{G}$ are two
non abelian isomorphic algebras which decompose into a direct sum of
algebras of general complex matrices.

If $h_{2}$ and $h_{1}$ are in relative generic position on $\mathcal{H}^{%
\mathbb{Q}}$, the previous analysis shows that $S$ and $S_{G}$ are
two isomorphic algebras. Moreover, these two algebras decompose into
a direct sum of one-dimensional quaternionic algebras.

On the contrary, if $h_{2}$ and $h_{1}$ are not in relative generic position
the previous analysis shows that $S$ and $S_{G}$ are two isomorphic algebras
which decompose into a direct sum of algebras of all quaternionic matrices.

So, we have found a correspondence between compatible quantum bi-Hamiltonian
systems in the Schroedinger and in the Heisenberg pictures, if the
associative product given in Eq.(\ref{associative}) coming from constant
K\"{a}hler metrics on the Hilbert space is assumed.

It is a simple matter to show that all these alternative associative
structures may be added to provide new alternative structures, i.e.
they are always compatible. Therefore the problem arises to find out
in which conditions we are going to find alternative associative
products whose classical limit would give the richness of
alternative Poisson structures of the classical situation.

Our belief is that the not necessarily compatible associative products found
in \cite{carinena} might correspond to alternative products not linearly
related when realized on the same manifold of states. In the coming section
we shall investigate these considerations by means of an example.

Moreover, the next example will show that choosing a non-constant
K\"{a}hler metric \cite{dub} changes the linear structure of the
space. In this case, different associative products between
operators, all compatible according to ref. \cite{carinena} have to
be introduced, in order to obtain the correspondence between
Heisenberg and Schroedinger pictures.

\subsection{Example: a two level system}

Here we will discuss the associative product between functions associated
with operators, but considering the manifold of states to be $\mathcal{M}%
\equiv \mathbb{R}^{4}$. A choice of a global chart on $\mathcal{M}$ allows
to write the tensors corresponding to an Hermitian tensor. The metric tensor
$G$ and the symplectic form $\Omega $ on $\mathbb{R}^{4}$ space will be
given by
\begin{equation}
G=dP_{1}\otimes dP_{1}+dQ_{1}\otimes dQ_{1}+dP_{2}\otimes
dP_{2}+dQ_{2}\otimes dQ_{2}
\end{equation}
and
\begin{equation}
\Omega =dP_{1}\wedge dQ_{1}+dP_{2}\wedge dQ_{2},
\end{equation}
so that the complex structure $J$ is
\begin{equation}
J=dP_{1}\otimes \frac{\partial }{\partial Q_{1}}-dQ_{1}\otimes \frac{%
\partial }{\partial P_{1}}+dP_{2}\otimes \frac{\partial }{\partial Q_{2}}%
-dQ_{2}\otimes \frac{\partial }{\partial P_{2}}.
\end{equation}
We may express them, recalling Eqs. (\ref{metrica}), (\ref{symp}) of the
previous simple example, in terms of different coordinates $%
q_{1},p_{1},q_{2},p_{2}$. Then\cite{bedlevo}
\begin{equation}
f_{A}(x)=\frac{1}{2}\langle x,Ax\rangle _{\mathcal{H}}=\frac{1}{2}\langle
x,A(G+i\Omega )x\rangle ,
\end{equation}
\begin{equation}
f_{2AB}(x)=\left[ f_{A}, f_{B}\right] _{\mathcal{H}}=\left( f_{A},%
f_{B}\right) _{g}+i\left\{ f_{A},f_{B}\right\} _{\omega }
\end{equation}
where $x=(q_{1}+ip_{1},q_{2}+ip_{2}),$ and
\begin{equation*}
\left( f_{A},f_{B}\right) _{g}=\sum\nolimits_{k}\left( \frac{%
\partial f_{A}}{\partial q_{k}}\frac{\partial f_{B}}{\partial q_{k}}+\frac{%
\partial f_{A}}{\partial p_{k}}\frac{\partial f_{B}}{\partial p_{k}}\right)
\end{equation*}
\begin{equation*}
\left\{ f_{A},f_{B}\right\} _{\omega }=\sum\nolimits_{k}\left( \frac{%
\partial f_{A}}{\partial q_{k}}\frac{\partial f_{B}}{\partial p_{k}}-\frac{%
\partial f_{A}}{\partial p_{k}}\frac{\partial f_{B}}{\partial q_{k}}\right) .
\end{equation*}

Denoting with $\sigma _{1}$, $\sigma _{2}$ and $\sigma _{3}$ the
realification of Pauli matrices on $\mathbb{R}^{4}$ space and with $\sigma
_{0}$ the identity matrix, in the coordinate system $Q_{1}$, $P_{1}$, $Q_{2}$%
, $P_{2}$, with $\{Q_{1},P_{1}\}=\{Q_{2},P_{2}\}=1,$ we have
\begin{equation*}
f_{\sigma _{0}}(x)=\frac{1}{2}\left[ Q_{1}^{2}+P_{1}^{2}+Q_{2}^{2}+P_{2}^{2}%
\right] ,
\end{equation*}
\begin{equation*}
f_{\sigma _{3}}(x)=\frac{1}{2}\left[ Q_{1}^{2}+P_{1}^{2}-Q_{2}^{2}-P_{2}^{2}%
\right] ,
\end{equation*}
\begin{equation*}
f_{2\sigma _{0}\sigma _{3}}(x)=\frac{1}{2}\langle x,2\sigma _{0}\sigma
_{3}(G+i\Omega )x\rangle =2f_{\sigma
_{3}}(x)=Q_{1}^{2}+P_{1}^{2}-Q_{2}^{2}-P_{2}^{2}
\end{equation*}
and
\begin{eqnarray*}
f_{2\sigma _{0}\sigma _{3}}(x) &=&\left( f_{\sigma _{0}},f_{\sigma
_{3}}\right) _{g}+i\left\{ f_{\sigma _{0}},f_{\sigma _{3}}\right\}
_{\omega } \\
&=&\sum\nolimits_{k}\left[ \frac{\partial f_{\sigma _{0}}}{\partial Q_{k}}%
\frac{\partial f_{\sigma _{3}}}{\partial Q_{k}}+\frac{\partial f_{\sigma
_{0}}}{\partial P_{k}}\frac{\partial f_{\sigma _{3}}}{\partial P_{k}}%
+i\left( \frac{\partial f_{\sigma _{0}}}{\partial Q_{k}}\frac{\partial
f_{\sigma _{3}}}{\partial P_{k}}-\frac{\partial f_{\sigma _{0}}}{\partial
P_{k}}\frac{\partial f_{\sigma _{3}}}{\partial Q_{k}}\right) \right]  \\
&=&Q_{1}^{2}+P_{1}^{2}-Q_{2}^{2}-P_{2}^{2}.
\end{eqnarray*}

On the contrary, it results
\begin{equation*}
\frac{1}{2}\langle x,2\sigma _{0}\sigma _{3}(G+i\Omega )x\rangle \neq \left(
f_{\sigma _{0}},f_{\sigma _{3}}\right) _{g}+i\left\{ f_{\sigma _{0}},%
f_{\sigma _{3}}\right\} _{\omega }
\end{equation*}
in the coordinates  $q_{1}, p_{1}, q_{2}, p_{2}$, when $\left\{
Q_{1},P_{1}\right\} =1/\left[ (1+\lambda
(q_{1}^{2}+3p_{1}^{2}))^{2}-4\lambda ^{2}p_{1}^{2}q_{1}^{2}\right] $ and $%
\left\{ Q_{2},P_{2}\right\} =1/\left[ (1+\lambda
(q_{2}^{2}+3p_{2}^{2}))^{2}-4\lambda ^{2}p_{2}^{2}q_{2}^{2}\right] .$

However, if we consider standard Poisson structures and Euclidean
structures in the $(q,p)$ variables, the quadratic functions
associated with Pauli matrices will define the expected product.

This example shows that we may realize the same abstract algebra in
two alternative ways not linearly related. To let them act on the
same manifold of states we have to endow this manifold $\mathcal{M}$
with two alternative linear structures (represented here by the
choice of $(q,p)$ and $(Q,P)$ variables respectively).

We believe that to realize these alternative algebras as algebras of
operators on the same vector space, this vector space must be
required to have much larger dimension (in the present case, it
should have a dimension greater than four). This situation should be
compared with the one of the (non-linear) Riccati equation and its
linearization as presented in Ref. \cite {lie}.

A realization of the Heisenberg algebra in terms of not linearly
related creation and annihilation operators has been given in Ref.
\cite {Wignerproblem} and considered to describe non-linear
oscillators.\cite {Physscriptaf-osc}

\subsection{A quaternionic example}

This example is provided to show that in the quaternionic setting further
problems arise with respect to the complex space situation already at the
level of ``constant'' tensor fields.

Let us consider a two level dynamical quantum system in the complex Hilbert
space $\mathcal{H}$ whose dynamics is described by the complex
anti-Hermitian time-dependent Hamiltonian $(\hbar =1)$
\begin{equation}
\widetilde{H}=2\Omega _{0}(t)J_{1}+2\Omega _{1}(t)J_{2}+\omega (t)J_{3},
\label{ham}
\end{equation}
where $\Omega _{0}(t)$, $\Omega _{1}(t)$ and $\omega (t)$ are real valued
functions of the time $t$ and the anti-Hermitian operators $J_{l}$ $(l=1,2,3)
$ obey the usual rules of commutation of the $su(2)$ algebra:
\begin{equation*}
\lbrack J_{l},J_{m}]=-\varepsilon _{lmn}J_{n}.
\end{equation*}

By resorting to the irreducible $2$-dimensional representation of the $J$
operators
\begin{equation}
J_{1}=\frac{i}{2}\left(
\begin{array}{cc}
0 & 1 \\
1 & 0
\end{array}
\right) ,\,\,\,\ J_{2}=\frac{1}{2}\left(
\begin{array}{cc}
0 & 1 \\
-1 & 0
\end{array}
\right) ,\,\,\,\ J_{3}=\frac{i}{2}\left(
\begin{array}{cc}
1 & 0 \\
0 & -1
\end{array}
\right)  \label{spin}
\end{equation}
and putting $\Omega =\Omega _{0}+i\Omega _{1}$, we can write the Hamiltonian
(\ref{ham}) as a $2\times 2$ anti-Hermitian complex matrix :
\begin{equation}
\widetilde{H}=i\left(
\begin{array}{cc}
\frac{\omega (t)}{2} & \Omega ^{\ast }(t) \\
\Omega (t) & -\frac{\omega (t)}{2}
\end{array}
\right) .\ \   \label{hamquat}
\end{equation}

The set $\mathfrak{H}$\ of anti-Hermitian complex operators obtained
by changing the entries in Eq.(\ref{hamquat}), is of course
irreducible in the $2$-dimensional Hilbert space $\mathcal{H}$,
since such is the spinorial representation (\ref{spin}) of the
$J_{l}$'s.

From a different point of view, we can interpret the Hamiltonian of Eq.(\ref
{hamquat}) as an anti-Hermitian quaternionic operator in a (right)
quaternionic Hilbert space $\mathcal{H}^{\mathbb{Q}}$, and the dynamics of
our quantum system is then described by the Schroedinger equation \cite{adl}
\begin{equation}
\frac{d}{dt}\left(
\begin{array}{c}
\Psi _{+}(t) \\
\Psi _{-}(t)
\end{array}
\right) =-i\left(
\begin{array}{cc}
\frac{\omega (t)}{2} & \Omega ^{\ast }(t) \\
\Omega (t) & -\frac{\omega (t)}{2}
\end{array}
\right) \left(
\begin{array}{c}
\Psi _{+}(t) \\
\Psi _{-}(t)
\end{array}
\right)
\end{equation}
where $\Psi _{+}(t),\Psi _{-}(t)\in \mathbb{Q}.$

Roughly speaking, $\mathcal{H}^{\mathbb{Q}}$ can be obtained from $\mathcal{H%
}$ by simply adding to each complex vector $\left| v\right\rangle
\in \mathcal{H}$ a term $\left| v^{\prime }\right\rangle j$, where
$\left| v^{\prime }\right\rangle \in \mathcal{H}$ and $j:j^{2}=-1$
is a quaternionic unity different from $i$; note that $\dim
\mathcal{H}^{\mathbb{Q}}=\dim \mathcal{H}=2$. \cite{sh}

The Cayley-Klein (CK) matrix reads \cite{dat, scolarici}
\begin{equation}
\left(
\begin{array}{c}
\Psi _{+}(t) \\
\Psi _{-}(t)
\end{array}
\right) =\left(
\begin{array}{cc}
F^{\ast } & L \\
-L^{\ast } & F
\end{array}
\right) \left(
\begin{array}{c}
\Psi _{+}(0) \\
\Psi _{-}(0)
\end{array}
\right) ,
\end{equation}
where $F(t)$ and $L(t)$ are complex functions depending on $\omega $ and $%
\Omega $ in a rather involved way; furthermore $F(0)=1,$ $L(0)=0$, and $%
|F|^{2}+|L|^{2}=1$ \cite{dat}.

The CK matrix can be regarded as the matrix representation of the time
evolution operator $U$ associated with the time dependent Hamiltonian (4),
and it belongs to a $2$-dimensional (complex) unitary representation of the $%
SU(2)$ group; by varying $H$ in $\mathfrak{H}$, we correspondingly obtain a
set $\mathfrak{U}=\{U\}$.

We remark once again that the form of any element $U\in \mathfrak{U}$ does
not depend on the scalar field, $\mathbb{C}$ or $\mathbb{Q}$, adopted. Now,
as long as we study the two-level system in $\mathcal{H}$, the set $%
\mathfrak{U}$ is clearly irreducible, hence, by the corollary of the Schur
Lemma, no non-trivial $G$ exists which commutes with it. Recalling the
discussion in sec. 3, we can conclude that the description of the system in $%
\mathcal{H}$ is unique.

On the contrary, if we now consider $\mathfrak{U}$ as a quaternionic group
representation acting on $\mathcal{H}^{\mathbb{Q}}$, it can be proven that
this representation is reducible into the direct sum of two equivalent
one-dimensional irreducible quaternionic representations on $\mathcal{H}^{%
\mathbb{Q}}$ \cite{fin2}, \cite{scoso}, so that $\mathfrak{U}$ admits a
non-trivial commutant. By a direct computation, the most general
quaternionic positive Hermitian matrix $G$ commuting with any $U\in %
\mathfrak{U}$ is
\begin{equation}
G=\left(
\begin{array}{cc}
a & jz \\
-jz & a
\end{array}
\right) ,\,\,\ z\in \mathbf{C},\,\,\ a>\left| z\right| .
\end{equation}

We can conclude that any element $U\in \mathfrak{U}$ is bi-unitary on $%
\mathcal{H}^{\mathbb{Q}}$ with respect to the Hermitian structures $%
h_{1}(\psi ,\varphi )=\langle \psi |\varphi \rangle _{1}$ and $h_{2}(\psi
,\varphi )=\langle \psi |\varphi \rangle _{2}$ with $\langle \psi |\varphi
\rangle _{2}=\langle \psi |G|\varphi \rangle _{1}$:
\begin{equation}
U^{\dagger }GU=G.
\end{equation}

Moreover, $h_{1}$ and $h_{2}$ are in generic position, in fact the
eigenvalues of $G$ are different, as one can prove by solving the eigenvalue
problem associated with it \cite{zha}: $\lambda _{1,2}=a\pm \left| z\right| $%
.

We show now that, according with results in the previous section,
the algebras $\mathfrak{U}\subset S=S_{G}$ decompose into a direct
sum of one-dimensional unimodular non-commutative algebras
$\mathbb{Q}$. In fact by applying the quaternionic unitary
transformation $D=\frac{1}{2}\left(
\begin{array}{cc}
1+i & j-k \\
1-i & j+k
\end{array}
\right) $, $D^{-1}=D^{\dagger }=\frac{1}{2}\left(
\begin{array}{cc}
1-i & 1+i \\
k-j & -j-k
\end{array}
\right) $ one obtains
\begin{equation*}
DU(t)D^{\dagger }=\left(
\begin{array}{cc}
F^{\ast }(t)-jL^{\ast }(t) & 0 \\
0 & F^{\ast }(t)-jL^{\ast }(t)
\end{array}
\right) ,
\end{equation*}
\begin{equation*}
DJ_{1}D^{\dagger }=\left(
\begin{array}{cc}
i & 0 \\
0 & i
\end{array}
\right) ,\ DJ_{2}D^{\dagger }=\left(
\begin{array}{cc}
j & 0 \\
0 & j
\end{array}
\right) ,\ DJ_{3}D^{\dagger }=\left(
\begin{array}{cc}
k & 0 \\
0 & k
\end{array}
\right)
\end{equation*}
and
\begin{equation*}
DGD^{\dagger }=\left(
\begin{array}{cc}
a+\left| z\right| & 0 \\
0 & a-\left| z\right|
\end{array}
\right) .
\end{equation*}

Finally, making resort to the form (8) of the evolution operator $U$, we can
also compute the transition probabilities in both the descriptions. Let us
for instance assume that the system is in the excited state $|+\rangle $ at $%
t=0$; the probability of finding the system in the ground state $|-\rangle $
at the time $t$ is given by
\begin{equation}
\mathcal{P}_{+\rightarrow -}(t)=|\langle -|U|+\rangle _{1}|^{2}=|L|^{2}
\end{equation}
according to the first description, and by
\begin{equation}
\mathcal{P}_{+\rightarrow -}^{\prime }(t)=|\langle -|U|+\rangle
_{2}|^{2}=|z|^{2}|F|^{2}+|L|^{2}
\end{equation}
according to the alternative description.

We emphasize in conclusion that the possibility of an alternative
description for this model can only occur in QQM, which then appears
as a theory intrinsically different from CQM, and not a mere
transcription of it. These findings seem to go in the same direction
as those found by Kossakowski in describing completely positive
maps. \cite{kossa}

\section{Conclusions}

In this paper, guided by the compatibility condition emerging for
Poisson structures when dealing with bi-Hamiltonian completely
integrable systems, we have considered the analog problem for
quantum systems in the framework of Schroedinger and Heisenberg
pictures. In particular, we have concentrated our attention on the
equivalence between the two descriptions when nonlinear
transformations on the manifold of states are performed. We find
that the two pictures are still equivalent when alternative
structures are taken into account without changing the linear
structure on the manifold of states. To allow for a nontrivial
compatibility condition on the associative structure of the
observables it seems that we are obliged to perform nonlinear
transformations. We have given an example where the mechanism is
present, however a reasonable understanding of the equivalence or
lack of it between the two pictures (Schroedinger and Heisenberg)
when nonlinear transformations are allowed is still missing and
further work is required.

\end{document}